\documentclass[traditabstract]{aa} 
\usepackage{graphicx}

\usepackage{natbib} 
\bibpunct{(}{)}{;}{a}{}{,} 
\usepackage{multirow,bigdelim}
\usepackage{rotating}
\usepackage[title]{appendix}
\usepackage{tabu} 

\usepackage{subfigure}
\usepackage{xcolor}

\newcommand{\virg}[1]{``#1"}
\def \ergsc{\hbox{erg s$^{-1}$ cm$^{-2}$}}

\def \fek {Fe K$\alpha$}
\def \chidof{$\frac{\chi ^{2}}{\rm d.o.f}$}

\begin{document}

\title{Prospects for the polarimetric mapping of the Sgr A molecular cloud complex with IXPE.}
\author{R. Ferrazzoli \inst{1,2,3}
 \and L. Di Gesu \inst{4}
 \and I. Donnarumma \inst{4}
 \and P. Soffitta \inst{1}
 \and E. Costa \inst{1}
 \and F. Muleri \inst{1}
\and  M. Pesce-Rollins \inst{5}
\and  F. Marin \inst{6}}
\institute
{INAF-IAPS, via del Fosso del Cavaliere 100, 00133, Roma, Italy
\and
Universit\`a di Roma \virg{Sapienza}, Dipartimento di Fisica, Piazzale Aldo Moro 5, 00185 Roma, Italy
\and 
Universit\`a di Roma Tor Vergata, Dipartimento di Fisica, via Cracovia 50 1,00133 Roma, Italy
\and
Agenzia Spaziale Italiana (ASI), Via del Politecnico snc, 00133, Roma, Italy
\and
Istituto Nazionale di Fisica Nucleare, Sezione di Pisa, I-56127 Pisa, Italy
\and
Universit\'e de Strasbourg, CNRS, Observatoire astronomique de Strasbourg, UMR 7550, F-67000 Strasbourg, France}
\date{}
\abstract{
	\textbf{Context.} 
	The X-ray polarization degree of the molecular clouds that surround Sgr A* is expected to be greatly lowered because the polarized reflection emission is mixed with the unpolarized thermal emission that pervades the Galactic center region. 
	For this reason, this observation is a challenging experiment for the upcoming Imaging X-ray Polarimeter Explorer (IXPE), whose launch is expected in 2021. 
	\\
	\textbf{Aims.}
	We aim to determine the detectability of four molecular clouds of the Sgr A complex (MC2, Bridge B2, Bridge E, and G0.11-0.11) in a realistic IXPE pointing of the Sgr A field of view.
	We assess the Minimum Detectable Polarization increase when a molecular cloud is off axis. 
	We provide two different strategies to reconstruct the intrinsic cloud polarization once the data will be available.
	\\
	\textbf{Method.}
	We use the Monte Carlo tool ixpeobssim to simulate IXPE observations of the Sgr A molecular cloud complex. 
	We use Chandra maps and spectra to model the diffuse emission in the Galactic center region along with a realistic model of the instrumental and diffuse background.  
	We create synthetic polarization products of the unpolarized emission.
	We combine them with a test dataset obtained from a simulation of a 2 Ms long IXPE observation to retrieve the intrinsic polarization degree of the molecular clouds.
	\\
	\textbf{Results.}
	We find that for the molecular clouds considered here, the MDP increases by $\sim$1$-$15\% with respect to the case in which a cloud is observed on-axis.
	We successfully retrieve the intrinsic polarization degree in the 4.0$-$8.0 keV band and line-of-sight distance of one of them taken as an example, G0.11-0.11, by correcting the observed (i.e., for a 2 Ms-long simulation) polarization degree map using either a synthetic dilution map or a Stokes intensity map of the unpolarized emission. 
	With both methods, the position of the cloud along the line-of-sight is derived from the reconstructed polarization degree with an uncertainty of 7 and 4 pc, respectively.
	\\
	\textbf{Conclusions.}
	We confirm the results of Di Gesu et al. (2020) that, assuming the distance along the line-of-sight and the polarization model of Marin et al. (2015), G0.11-0.11 is the most promising target. 
	For the Sgr A molecular complex region we propose an observation strategy that may permit to detect up to three clouds in the 4.0$-$8.0 keV band depending on their true line-of-sight position.
	We demonstrate that, by using simulated data products of the unpolarized components, it is possible to clean-up the observed polarization maps from the environmental contamination. 
	The methods that we present here are potentially useful for the analysis of X-ray polarimetric data of any extended source that is affected by environmental dilution of the polarized signal.
	To measure accurately (i.e with uncertainties of the order of a few parsec) the distance of the cloud along the line-of-sight, a high-quality spectrum and image of the clouds quasi simultaneous to the IXPE pointing are needed.
}
\keywords{Polarization, Galaxy:nucleus, X-rays:general}
\maketitle
\titlerunning{Prospects for the polarimetric mapping of the Sgr A molecular cloud complex with IXPE}
\authorrunning{R. Ferrazzoli et al.}

%

\section{Introduction}
\label{intro}
\begin{figure}[htbp]
	\centering
	\includegraphics[width=0.5\textwidth]{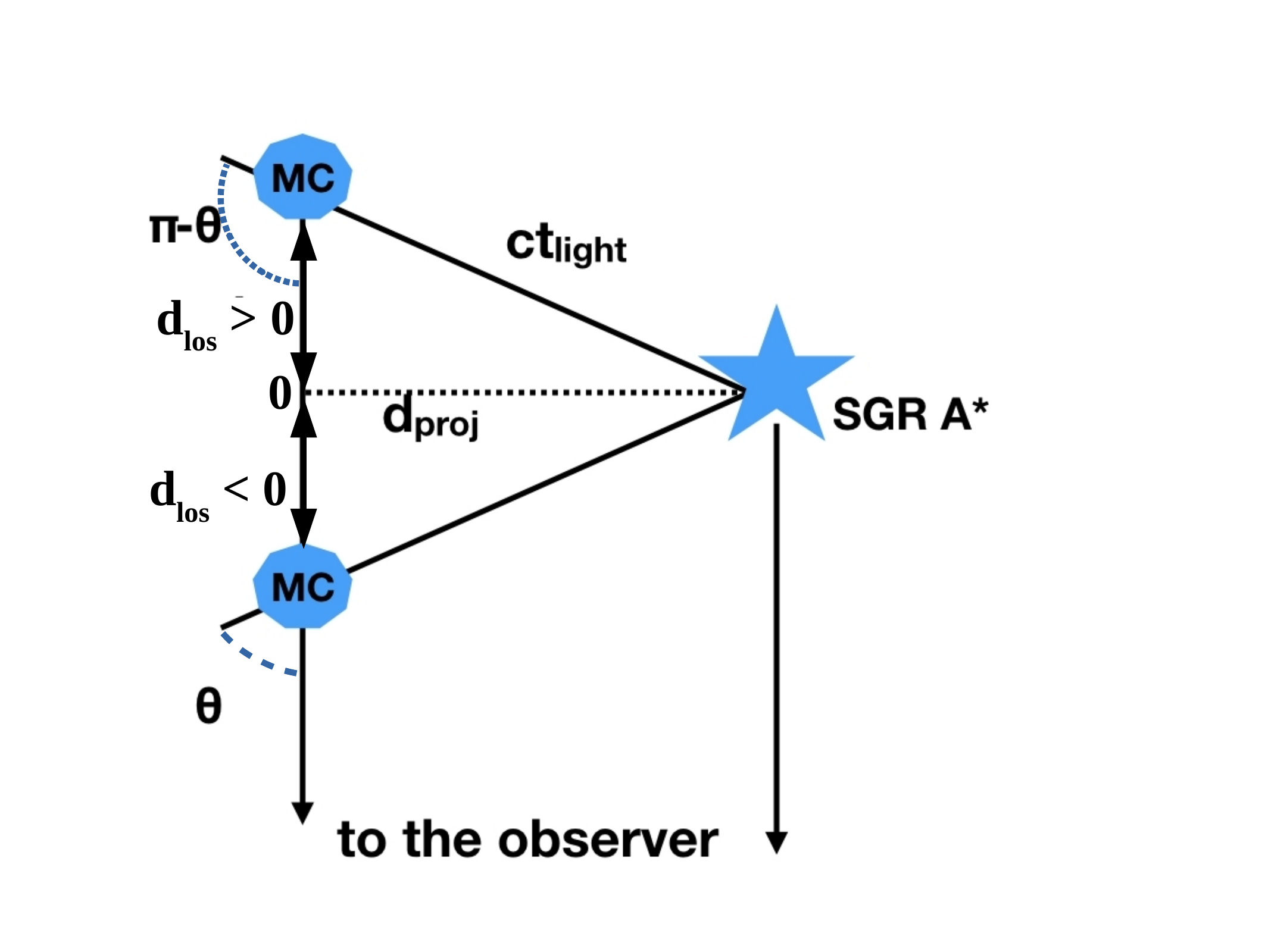}
	\caption{Scattering geometry for a MC located in front or behind the Sgr A* plane. 
		The two positions depicted have scattering angles $\it \theta$ and $\it \pi-\theta$ which results in the same polarization degree. 
		In addition, $\it \vec{d}_{\rm proj}$ is the cloud-Sgr A* distance projected in the plane of the sky, $\it\vec{d}_{\rm los}$ is the line-of-sight distance of the cloud with respect to the Sgr A* plane, $\it c$ is the speed of light, and $\it t_{\rm light}$ is light travel time between Sgr A* and the cloud.
	The vector $\it\vec{d}_{\rm los}$ assumes negative values if the cloud is in front of the Sgr A* plane and positive if behind.}
	\label{fig:figure_1}
\end{figure}

Determining the luminosity history of Sgr A*, the supermassive black hole (SMBH) that lies in the center of our Galaxy, would be of great interest for our understanding of the duty cycle of mass accretion onto SMBHs, which is thought to drive the coevolution of SMBHs and galaxies \citep{dimatteo2008}.
Many phenomena observed in the Galactic center (GC) region point to a past activity of  Sgr A* \citep[see][for a review]{ponti2013}.
For instance the gamma and X-ray bubbles observed by Fermi-LAT and eROSITA above and below the Galactic plane are indicative of an Active Galactic Nucleus (AGN) phase of Sgr A* some million years ago \citep{su2010, zubovas2011, Predehl2020}. \\
In the last 30 years, X-ray spectral \citep[e.g.][]{Sunyaev1993, Koyama1996, Murakami2000, ryu2013,capelli2012, walls2016, chuard2018} and timing \citep[e.g.][]{muno2007,Inui2009,ponti2010,clavel2013, terrier2018} studies of the molecular clouds (MC, such as MC2, Bridge B2, G0.11-0.11, Bridge E, Sgr B2, Sgr C1, Sgr C2, and Sgr C3) that are located in the 100 pc region surrounding Sgr A* have provided evidence of past single \citep{ponti2010} or multiple \citep{clavel2013, terrier2018} outburst of Sgr A*. 
Indeed, the clouds display X-ray reflection spectral features like a steep continuum plus a \fek \,  emission line that are variable in time.
However, no possible X-ray bright illuminating source is present nearby, which led \citet{Sunyaev1993} to suggest that the observed X-ray reflection spectrum from the MC is the echo of a past outburst of Sgr A*, delayed by the light travel time across the Central Molecular Zone.
This hypothesis implies that some hundred years ago Sgr A* was $10^6$ times more luminous than today and similar to a low-luminosity AGN.
Other possible sources of illumination of the clouds include cosmic rays (LECR) from a local source \citep{yusef-zadeh2013,dogiel2014}. 
For example, \citet{zhang2015} and \citet{mori2015} note that there is still not enough evidence to completely exclude that CR are responsible for at least part of the GC X-ray steady emission. \\
Despite many observational efforts, it is still difficult to unambiguously derive the past light curve of Sgr A* from the X-ray variability of the MC. 
This is mainly because the distance $\it \vec{d}_{\rm los}$ of the clouds along the line-of-sight is loosely constrained \citep[e.g.,][]{Reid2009,capelli2012,walls2016, chuard2018}, which, in turn, makes it challenging to accurately infer the light travel time $\it t_{\rm light}$. 
We sketch the geometry of the MC-SgrA* system in Fig. \ref{fig:figure_1}, where $\it \vec{d}_{\rm proj}$ is the the Sgr A*-cloud distance projected on the plane of the sky, $c$ is the speed of light and $\theta$ is the scattering angle. \\
A possible way to overcome the difficulty in determining $\it \vec{d}_{\rm los}$ is provided by X-ray polarimetry.
If the MC were illuminated by an external compact source like Sgr A*, the reflected X-ray radiation would be highly linearly polarized by scattering. 
The expected polarization degree $\it P$ depends on the scattering angle $\it \theta$ as:
\begin{equation}
	\rm	P = \frac{1 - \cos^2\theta}{1 + \cos^2\theta} \quad .
	\label{eq:poldeg}
\end{equation}
In turn, the scattering angle is related to $\it \vec{d}_{\rm los}$ by:
\begin{equation}
	\rm	\vec{d}_{los} = \vec{d}_{proj}\cot\theta \quad .
	\label{eq:distance}
\end{equation}
In this scenario, the polarization degree is 100\% for a cloud located in the Sgr A* plane ($\it \vec{d}_{\rm los}=\rm 0$ pc, $\it \theta=\rm 90\degr$), while the direction to the external illuminating source is perpendicular to the polarization direction \citep{Vainshtein1980}.
Therefore, detecting the polarization degree of the molecular clouds would identify the location of the illuminating source and produce a map of the molecular clouds in the GC region in three dimensions. 
The two cloud positions shown in Fig. \ref{fig:figure_1} result in the same polarization.
However, this degeneracy can be broken making use of spectral information, because the shape of the reflected continuum at low energies also depends on the scattering angle \citep{churazov2002}. \\
Thanks to the imminent launch of the Imaging X-ray Polarimetry Explorer (IXPE, \citealt{weisskopf2016}) in late 2021, it will be possible, for the first time, to employ spatially resolved X-ray polarimetry to address the Sgr A* past outburst hypothesis in an independent way. \\
The prospect of having soon an X-ray polarimeter has led to a renewed interest in the modeling of the X-ray polarization properties in the GC region \citep{churazov2002,churazov2017,churazov2017-2,Khabibullin2020} and in evaluating the detectability of candidate molecular cloud targets \citep{marin2014,marin2015}. \\
On the observational side, in \citet{DiGesu2020} we set up a method to perform realistic simulations of IXPE observations of the MC and of their environment considering the polarimetric, spatial, and spectral properties of all the components that contribute to the X-ray emission in the GC region. 
Indeed, besides the clouds, there are two thermal components that contribute to the 2$-$8 keV emission in the GC: a $\sim$1 keV soft plasma, and a thermal component that is often modeled as a 6.5 keV hard plasma \citep[e.g.,][]{ryu2013}. 
These thermal components permeate the GC region and are unpolarized. 
Thus, the detected polarization degree of the MC is lower than the intrinsic value by a factor that depends mainly on the amount of plasma contamination in the surrounding environment.
For instance, in \citet{DiGesu2020} we found a diluting effect of the plasma as high as 90\% in the 2$-$4 keV band, and 60\% in the 4$-$8 keV band. \\
In this work, we expand upon \citet{DiGesu2020} by simulating a long-lasting IXPE observation of the entire Sgr A field of view (FOV), rather than individual clouds on axis. 
Indeed, a single IXPE pointing of the Sgr A complex will capture more than one cloud at different off-axis positions.
It is therefore relevant to address the issue of how the detectability changes when a cloud is not at center of the FOV. 
In addition, our simulation method has the advantage of treating all the components that contribute to GC emission separately, each one with its own spectral and morphological property, that are well known thanks to the legacy of Chandra.
It is reasonable to assume that the diffuse plasma in the GC does not change in spectrum and morphology over time.
This implies that it is possible to exploit our simulations to create synthetic products of the diluting components with the aim of combining them with real data to recover the undiluted polarization degree of the MC.
In this work, we test two methods to achieve this goal. \\
In the following, we use the Stokes parameters formalism to describe the polarization properties \citep{Stokes}.
The Stokes parameters describe the polarization degree $\it P$ as the fraction between polarized and unpolarized flux:
\begin{equation}
\rm P = \frac{\sqrt{Q^2 + U^2}}{I} \quad ,
\label{eq:P_stokes}
\end{equation}
where $\it I$ is the total intensity,  $\it Q$ represents the linearly polarized radiation intensity along the reference frame axes, and $\it U$ is the linearly polarized radiation intensity at $\pm$45\degr with respect to the main reference frame axis. \\
Throughout the paper we quantify the detectability of the targets by computing the  Minimum Detectable Polarization (MDP, \citealt{weisskopf2010}).
The MDP is a fundamental quantity for the statistical significance of an X-ray polarization measurement that represents the degree of polarization which can be determined with a 99\% probability against the null hypothesis, and is defined as:
\begin{equation}
	\rm MDP = \frac{4.29}{\mu R_S} \sqrt{\frac{R_S + R_B}{T}}
	\label{eq:mdp}
\end{equation}
where $\it R_{\rm S}$ is the detected source rate (in counts/s), $\it R_{\rm B}$ is the background rate (in counts/s), $\it T$ is the observation time (in seconds) and $\it \mu$ is the adimensional modulation factor of the detector, i.e., the response of the detector to a 100\% polarized radiation at a given energy. \\
The paper is organized as follows: in Sect. \ref{sec:methods} we describe the setup of our simulations, in Sect. \ref{sec:detectability} we present a simulated MDP map to identify the detectable targets, and we discuss how the detectability changes with the position of the targets in the FOV. 
In Sect. \ref{sec:reconstruction} we present two methods to recover the intrinsic polarization degree of the MC using synthetic products of the diluting components. 
Finally, we discuss our findings in Sect. \ref{sec:discussion} and we outline our conclusions in Sect. \ref{sec:conclusion}.
\begin{figure*}[htbp]
	\centering
	\includegraphics[width=0.85\linewidth]{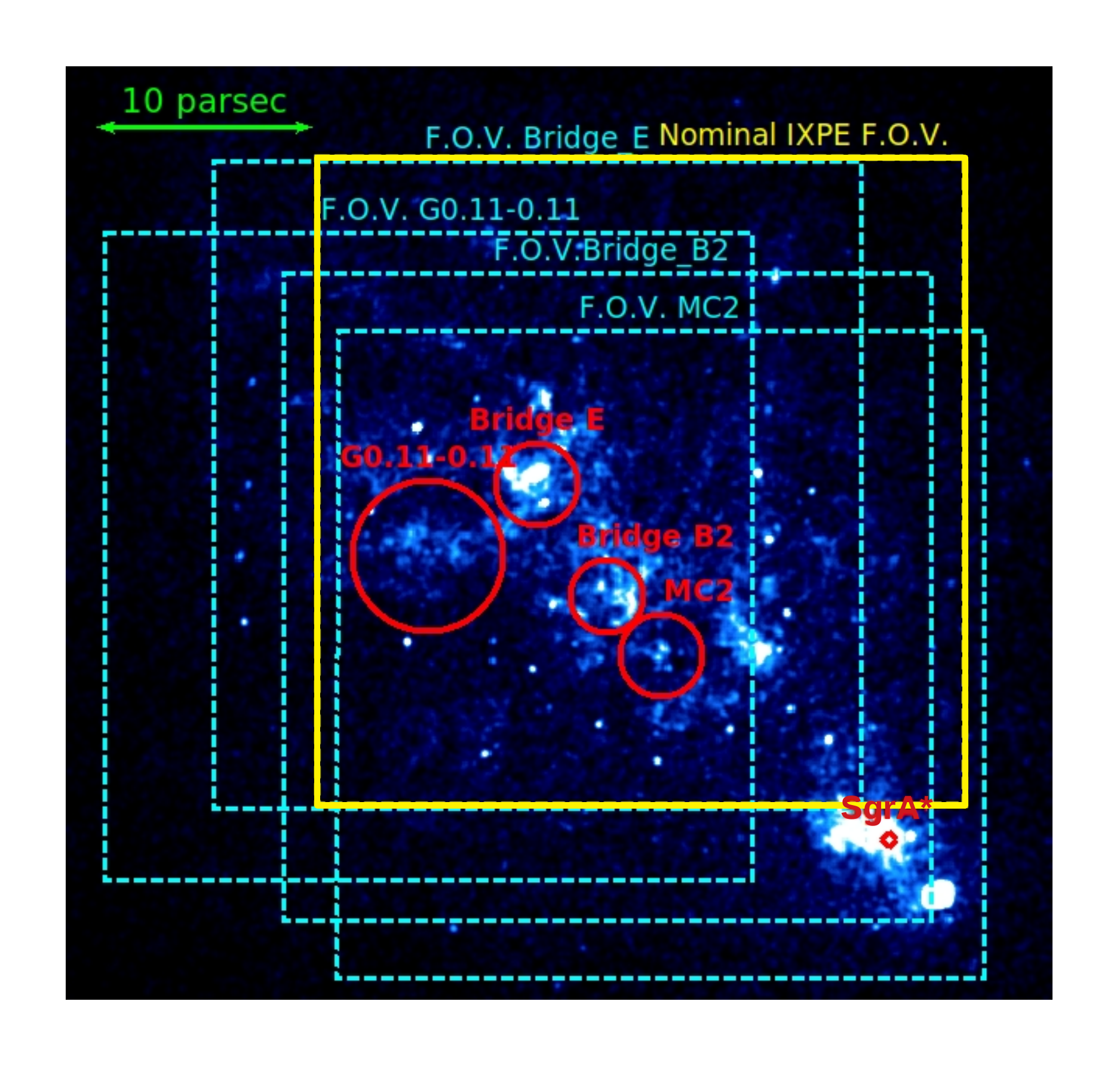}
	\caption{Chandra \fek \, map of the Sgr A complex. 
		The dashed boxes are the FOVs considered in the analysis of Sect. \ref{sec:offaxis}, while the solid box represent the nominal IXPE FOV of our baseline simulation.
		The circles display the clouds considered in this work, while the diamond marks the position of Sgr A*.
		The double-headed arrow represents a distance of 10 pc.}
	\label{fig:figure_2}
\end{figure*}
%
%
\section{Method}
\label{sec:methods}
\subsection{Source model}
\label{sec:sourcemodel}
\begin{figure*}[htbp]
	\centering
	\includegraphics[width=1\linewidth]{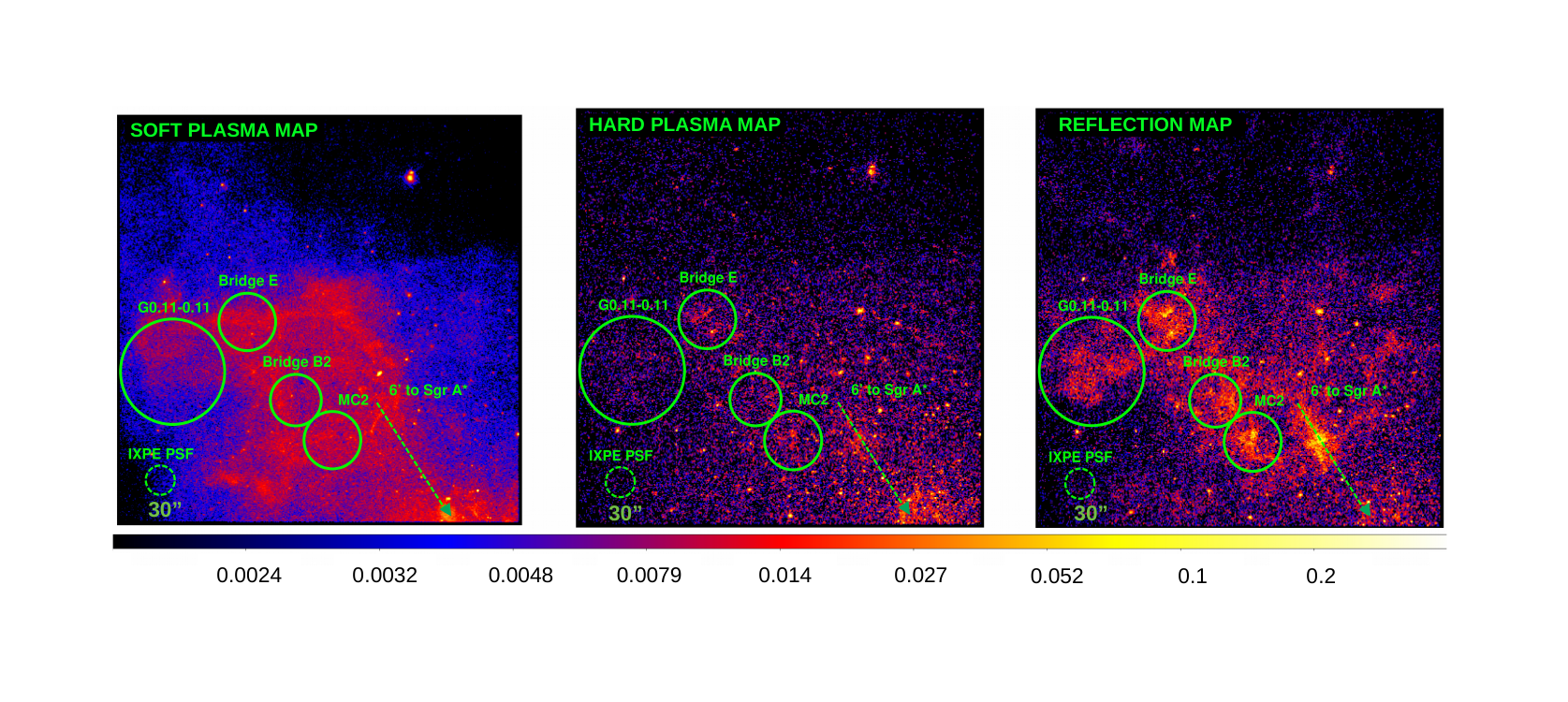}
	\caption{From left to right: background and continuum-subtracted merged Chandra maps of the soft plasma, the hard plasma, and the reflection components in the Sgr A MC complex region centered on the nominal IXPE pointing. 
		The images are smoothed using a 3 pixel Gaussian kernel.
		The color bar displayed on the bottom has adimensional units because the images are normalized to the maximum value.
		The regions shown in the solid circles are the MC considered for IXPE simulations (i.e., MC2, Bridge B2, Bridge E, and G0.11-0.11).
		A dashed circle having the size of the IXPE PSF is shown for comparison. 
		The direction to Sgr A* is indicated with a dashed arrow.}
	\label{fig:figure_3}
\end{figure*}

\begin{figure*}[htbp]
	\centering
	\includegraphics[width=1\linewidth]{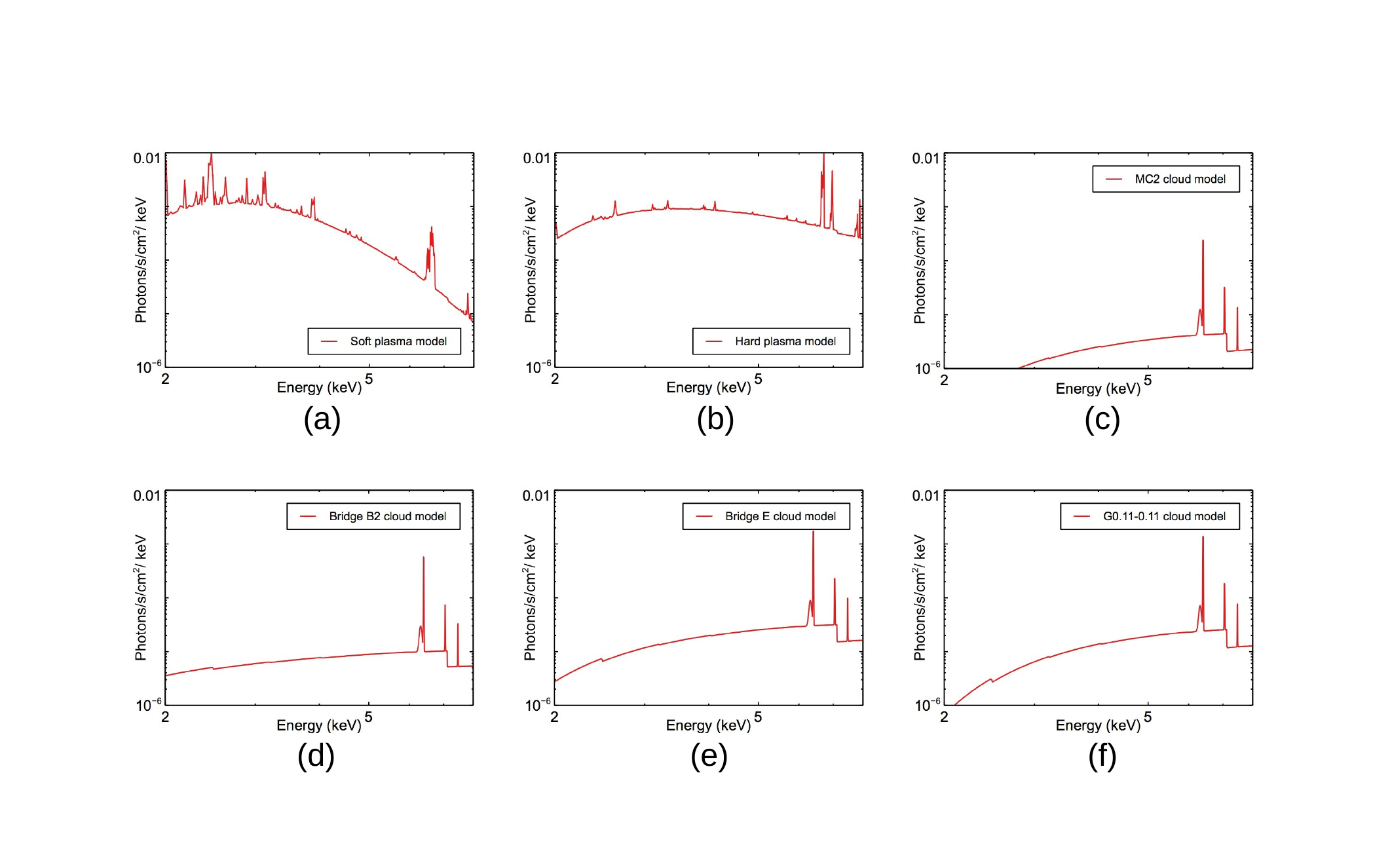}
	\caption{Spectral models for the emission components of our simulations: the soft (a) and the hard (b) plasma in the nominal FOV, and the reflection in the MC2 (c), Bridge B2 (d), Bridge E (e), and G0.11-0.11 (f) region.
		The models were obtained from the Chandra spectral analysis performed in the present work and in \citet{DiGesu2020}}.
	\label{fig:figure_4}
\end{figure*}

In this work, we simulate IXPE observations of the Sgr A molecular complex and we investigate the detectability of the MC MC2, Bridge B2, G0.11-0.11, and Bridge E. 
As in \citet{DiGesu2020}, we do not consider the MC Bridge D and MC1 because they are expected to be basically unpolarized according to the model of \citet{marin2015}. \\
Throughout the paper we use as a main pointing the 12.8'$\times$12.8' IXPE FOV centered on coordinates fk5 17:46:02.4020, -28:53:23.981, shown in Fig. \ref{fig:figure_2}. 
This position is centered on the X-ray reflection feature known as \virg{the bridge} that is considered one of the most promising targets for X-ray polarimetric observations because its average emission has been persistently bright in the last ten years \citep{churazov2017}. 
This is the region of interest of our baseline simulation.
Hence, in Sect. \ref{sec:offaxis}, we test other possible pointings centered on each MC to see how the detectability changes with the location of the target in the FOV. \\
Following \citet{DiGesu2020}, in the region(s) of interest we simulate all the diffuse components that contribute to the emission in the GC region. 
The soft and hard plasma components are simulated over the entire FOV. 
In order to account for the morphology of the plasma, we created background and continuum-subtracted Chandra maps.
For the soft plasma, we used the 1.7$-$3.3 keV energy band that comprises the \ion{S}{xv} and \ion{Ar}{xvii} emission lines, while for the hard plasma we created a map centered on the energy of the \ion{Fe}{xxv}-He$\alpha$ emission line (6.62 $-$ 6.78 keV). 
We created the maps using the procedure outlined in \citet{DiGesu2020} to combine 2.4 Ms of archival Chandra-ACIS data.
We show the soft and hard plasma maps in the first and second panel of Fig. \ref{fig:figure_3}.
We extracted the spectrum of the plasma components for all the IXPE FOV from the latest available Chandra observation that contains the IXPE nominal pointing (i.e., Chandra OBS ID 20808 from 8 October 2017). 
After subtracting the blanksky and removing the point sources, that we identified through the CIAO tool wavdetect with 2 and 4 pixel scales and 10$^{-6}$ signal threshold, we extracted the spectra over the whole FOV centered at the nominal pointing. 
We note that, in the regions that we used for the MC, there are no point sources \citep{DiGesu2020}.
Thus, there is no need to remove the points sources from the Chandra maps because they have no impact for our regions of interest.
A transient appearing by chance in our region of interest during the IXPE observation should have a flux above 4$\times$10$^{-13}$ erg/s/cm$^2$ (i.e the uncertainty on the total flux of G0.11.011 see Table \ref{tab:clouds}) to cause a sensible contamination. 
In a real observation, the transient can be removed either by cutting a PSF-large region from the maps or by removing the contaminated time intervals from the event files. \\
We fitted the spectrum with XSPEC \citep[][version 12.10.1]{xspec} in the 2.0$-$8.0 keV band obtaining a reduced \chidof $\sim 1.5$.
We model the Galactic absorption with the phabs model. 
We fitted the plasma components with a collisionally-ionized plasma model \citep[APEC,][]{smith2001} with a temperature set to 1.0 keV for the soft plasma and 6.5 keV for the hard plasma, and solar abundances. 
For the reflection component, we used the neutral reflection model PEXMON \citep{nandra2007} that consistently models both the continuum and the \fek\, emission.
These spectral models are commonly used for fitting the GC diffuse X-ray emission \citep[see e.g. ][]{2009Ryu,ponti2010,ryu2013,mori2015}. 
The model spectra of the plasma derived from this fit (first and second panel of Fig. \ref{fig:figure_4}) serve as input in our simulations.
When simulating other pointings, we extract the spectrum again to match the new coordinates. 
We note that the flux of the plasma does not change significantly from a pointing to another. 
We consider all the plasma components as unpolarized. \\
The reflection component of the MC is simulated over circular regions as listed in Table \ref{tab:clouds}.
For their morphology and spectral properties, we use the same Chandra maps and spectra of \citet{DiGesu2020}. 
These are continuum and background subtracted Chandra maps centered on the Fe-K$\alpha$ line (6.32$-$6.48 keV) and cut over circular regions having the radius of the cloud listed in Table \ref{tab:clouds}.
In the third panel of Fig. \ref{fig:figure_3} the cloud regions are shown superimposed to the \fek \, map of the whole FOV.
The model spectra of each MC is shown in Fig. \ref{fig:figure_4}.
We take the polarization properties from the modeling of \citet{marin2015}. 
We consider the polarization degree of the reflection component as constant with energy, but null at the energy of the fluorescence Fe-K$\alpha$ line, because the fluorescent lines from spherically symmetrical orbitals are unpolarized. 
In Table \ref{tab:clouds} we list, for each cloud region and for the entire FOV, the polarization properties of all the spectral components and the flux contributions in each region. 
The polarization degree values that we assume in our simulations were derived in \citet{marin2015} assuming the distance along the line-of-sight $\it d_{\rm los}$ that are listed in Table \ref{tab:clouds}. 
In addition, we list in Table \ref{tab:clouds} other possible values of $\it \vec{d}_{\rm los}$ \citep{capelli2012} and the correspondent polarization degree resulting from Eq.\ref{eq:poldeg} and \ref{eq:distance}. 
The ranges of distances calculated by \citet{capelli2012} include $\it d_{\rm los}=0$.
Thus, they represent an upper limit for the absolute value of the distance along the line-of-sight, from which Eq. \ref{eq:poldeg} and \ref{eq:distance} returns the maximum theoretical polarization degree of the clouds.
Because the value of the theoretical polarization degree depends strongly on the assumption of $\it \vec{d}_{\rm los}$, we consider also these alternative polarization degree values in the discussion of the cloud detectability in Sect. \ref{sec:detectability}. 
Finally, we include in our model the Cosmic X-ray background (CXB) and the IXPE instrumental background.
The CXB is simulated as a uniform source over the entire FOV with the spectrum of \citet{moretti2009}.
In \citet{DiGesu2020} the instrumental background was based on the one measured for the Neon filled detector on board of the OSO-8 experiment \citep{bunner1987}.
We now employ a realistic instrumental background spectrum that is based on the estimates of \citet{xie2021}.
They found for the IXPE detector a background level of $1.16 \times 10^{-2}$ counts s$^{-1}$ cm${^2}$ in 2$-$8 keV. 
We discuss the effect and removal of the instrumental background in the Appendix \ref{sec:appendix}.
%
\subsection{Simulation outputs}
\label{sec:simulation}
We simulate IXPE observations of the Sgr A MC complex with the Python-based framework ixpeobssim 12.0.0. \citep{pesce2019}.
The framework convolves the user defined source model, including morphological, spectral, and polarization properties, with the IXPE instrument response functions (i.e., the PSF, the telescope effective area, and the vignetting) to produce the simulated event files.
These can be used to create images, spectra, and maps of the Stokes parameters with user defined spatial and energy binning. \\
In the following analysis, we make use of MDP map cubes and polarization map cubes. 
The MDP map cubes are data structures that contain the information needed for the calculation of the MDP (i.e., mean energy, counts, effective modulation factor) binned in sky coordinates for each energy bin considered.
We employ them to produce the MDP maps (see Sect. \ref{sec:mdpmap}). 
Conversely, the polarization map cubes hold polarization information binned in sky coordinates and contain image extension for the Stokes parameters $\it I$, $\it U$, $\it Q$, and for the polarization degree and angle. 
We use the Stokes parameters map contained in the polarization map cubes for the retrieval of the polarization degree of the MC (see Sect. \ref{sec:dilution} and \ref{sec:subtraction}). \\
To mimic a real IXPE observation (Sect. \ref{sec:mdpmap}, \ref{sec:offaxis}, \ref{sec:dilution}, \ref{sec:subtraction}) we run simulations with exposure time of 2 Ms (that hereafter we label as \virg{realistic} simulations), which is a realistic estimation of the time that IXPE will dedicate to the observation of the GC during the first two years of operations.
In order to create synthetic polarization products accounting for the effects of the unpolarized components (Sect. \ref{sec:dilution}, \ref{sec:subtraction}), we run simulations of 200 Ms (that hereafter we label as \virg{ideal} simulations) that reaches a MDP of less than 1\% over the entire FOV. 
This long exposure time serves to minimize the statistical uncertainty (error on $\it P \ll \rm 1\%$) of the result of the simulation.
With this simulation setup, we convert the model of the unpolarized components into IXPE data products without adding uncertainty. 
Thus, the synthetic maps are affected only by the uncertainty that derives from the spectral fit of the Chandra data on which the input model is based  (Sect. \ref{sec:sourcemodel}).
\begin{table*} \tiny
\caption{Input data for IXPE simulations of the nominal FOV.}     
\label{tab:clouds}      
\centering         
{\tabulinesep=1.mm           
\begin{tabu}{lcccccccc}        
\hline\hline                 
Region\tablefootmark{a}	&\multicolumn{2}{c}{Region \tablefootmark{b}}& $\rm \vec{d}_{proj}$\tablefootmark{c}& $\rm \vec{d}_{los}$\tablefootmark{d}& $\rm \vec{d}_{los}^{other}$\tablefootmark{e} & $\rm P_{model}$\tablefootmark{f} &$\rm P_{other}$\tablefootmark{g} & Model component fluxes\tablefootmark{h}\\
&\, \, \, \, center		& size				&		&  		&				&		&			& Soft plasma: 4.0-8.0 keV\\
&   					&  					& 		& 		& 				&		& 			& Hard plasma: 4.0-8.0 keV\\
& 			 			&  					& 		& 		&  				& 		& 			& Reflection continuum: 4.0-8.0 keV\\
& (hh:mm:ss.s, 			&  					& 		& 		&  				& 		& 			& \fek: 4.0-8.0 keV\\
& dd:mm:ss.s) 			& \arcmin 			& (pc)	& (pc)	& (pc)  		& (\%) 	& (\%) 		&($10^{-12}$ \ergsc) \\
\hline
\multirow{4}{*}{MC2}	& circle: 			& \multirow{4}{*}{(radius) \, 0.82}& \multirow{4}{*}{-14}& \multirow{4}{*}{-17}	& \multirow{4}{*}{-29.7$-$7.3}& \multirow{4}{*}{25.8} 	& \multirow{4}{*}{$\ge$10} 												& $5 \pm 4$\\
& (17:46:00.6,			&  	 				&	 	& 		&  				& 		& 			& $3.9\pm 0.8$  \\
& -28:56:49.2)			&					&		&		&				&		&			& $0.2 \pm 0.1$\\
&						&					&		&		&				&		&			& $1.7 \pm 0.7$\\
\hline
\multirow{4}{*}{Bridge B2}	& circle:& \multirow{4}{*}{(radius) \, 0.73}	& \multirow{4}{*}{-18}	& \multirow{4}{*}{-60}		& \multirow{4}{*}{-6.9$-$6.9}	& \multirow{4}{*}{15.8} 		& \multirow{4}{*}{$\ge$77.3}										& $0.3 \pm 0.2$ \\
& (17:46:05.5,			&  	& 				& 		&  		&  				& 					& $1.9 \pm 0.6$\\ 
& -28:55:40.8) 			&					&		&		&				&		&			& $0.4 \pm 0.1$\\
&						&					&		&		&				&		&			& $4.3 \pm 0.7$\\
\hline
\multirow{4}{*}{G0.11-0.11}	& circle: & \multirow{4}{*}{(radius) \, \, 1.5} 	&  \multirow{4}{*}{-27} &  \multirow{4}{*}{-17} &  \multirow{4}{*}{-3.1$-$3.1}	&  \multirow{4}{*}{55.8} 		&  \multirow{4}{*}{$\ge$62.5} 										& $3.2 \pm 0.4$\\
& (17:46:21.6,	& 			& 	& 	& 	& 	& 													& $9 \pm 1$ \\
& -28:54:52.1) 	&					&		&		&				&		&					& $1.0 \pm 0.1$\\
&				&					&		&		&				&		&					& $10.0 \pm 0.9$\\
\hline
\multirow{4}{*}{Bridge E}& circle:	& \multirow{4}{*}{(radius) \, 0.82} & \multirow{4}{*}{-25}	& \multirow{4}{*}{-60}	&  \multirow{4}{*}{-13.7$-$13.7}	& \multirow{4}{*}{12.7}		& \multirow{4}{*}{$\ge$97.4}										& $0.5 \pm 0.2$\\
&  (17:46:12.1,	&  			& 			& 	& 	&  	& 											& $4.7 \pm 0.8$\\
& -28:53:20.3) 	&					&		&		&				&		&					& $1.3 \pm 0.1$\\
&				&					&		&		&				&		&					& $12.7 \pm 0.9$\\
\hline
\multirow{4}{*}{Field of view} 	& box: 				& \multirow{4}{*}{(side) \, \, \,12.8}		& \multirow{4}{*}{-} 		&  \multirow{4}{*}{-} 		&  \multirow{4}{*}{-} 				&  \multirow{4}{*}{-} 		&  \multirow{4}{*}{-} 			& $18.9 \pm 1.4$ \\
& (17:46:02.4,					&					& 		& 		& 				& 		& 			& $71.3 \pm 1.4$ 	\\
& -28:53:23.98)					& 					&  		&  		&  				&  		&  			& - \\
& 								& 					&  		&  		&  				&  		&  			& - \\
\hline	\hline
\end{tabu}}
\tablefoot{
\tablefoottext{a}{Region name. 
	The MC are cross identified with the targets listed in \citet{marin2015}.}
\tablefoottext{b}{Geometrical dimensions of the region over which each spectral components is simulated, as described in Sect. \ref{sec:methods}.} 
\tablefoottext{c}{Projected distance from Sgr A*. Negative values: MC East of the GC.} 	
\tablefoottext{d}{Line-of-sight distance $\it \vec{d}_{\rm los}$ from \citet{marin2015}. Negative values: MC in front of the Galactic plane.}
\tablefoottext{e}{Range of other line-of-sight distances $\it \vec{d}_{\rm los}$ from \citet{capelli2012}.
	Because these ranges of distance include $\it d_{\rm los}=0$, they prescribe upper limits for the maximum theoretical polarization degree of the clouds.}
\tablefoottext{f}{Polarization degree $\it P$ from \citet{marin2015}.}
\tablefoottext{g}{Range of polarization degree correspondent to  $\vec{d}_{\rm los}^{\rm other}$, obtained from Eq. \ref{eq:poldeg} and \ref{eq:distance}}.
\tablefoottext{h}{Flux contribution of each spectral component in the given region. 
	Fluxes are taken from the spectral fits of \citet{DiGesu2020} and Sect. \ref{sec:methods}. }
}
\end{table*}
\section{Target detectability}
\label{sec:detectability}
\subsection{MDP map}
\label{sec:mdpmap}
\begin{figure*}[htbp]
	\centering
	\includegraphics[width=0.8\textwidth]{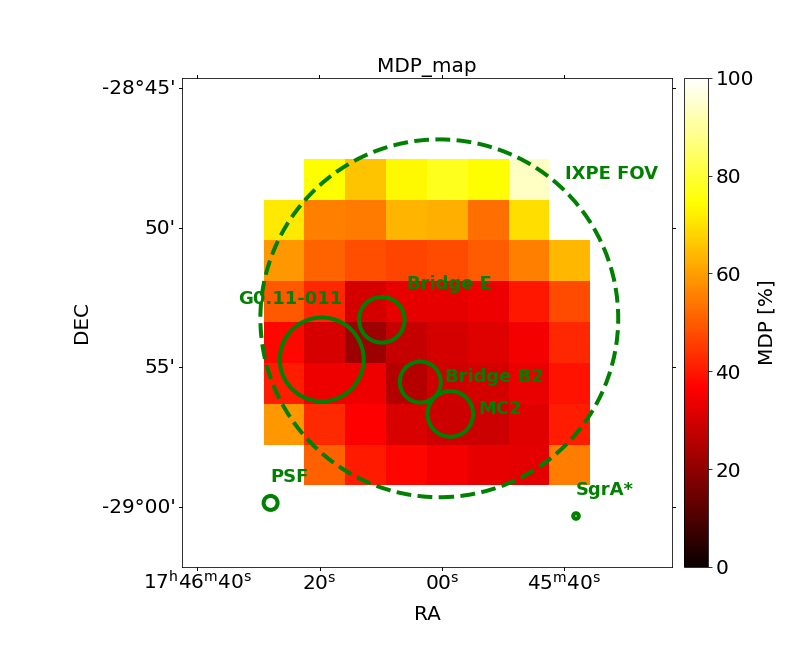} 
	\caption{MDP map with 1.5\arcmin \, spatial binning and 2 Ms exposure in the 4$-$8 keV energy band.
			The dashed circle represents the IXPE FOV. 
			The solid circles are the MC regions considered for IXPE simulations (i.e., MC2, Bridge-B2, Bridge-E, and G0.11-0.11).
			Smaller circles indicate the size of the IXPE PSF and the position of Sgr A*.}
	\label{fig:figure_5}
\end{figure*}

We created the MDP map for a 2 Ms-long IXPE observation centered on the nominal pointing.
The MDP map allows us to identify the regions for which the MDP reaches the lowest value.
In Fig. \ref{fig:figure_5}, we show the MDP map in the 
4$-$8 keV energy band, where the polarized reflection component outshines the plasma emission.
The maps are obtained from the MDP map cubes described in Sect. \ref{sec:simulation}.
We bin the map with a sky pixel size of $\sim$1.5'.
This corresponds to three times the IXPE PSF and to the typical size of the MC (Table \ref{tab:clouds}).
Minima of MDP are observed in the region of the MC G0.11-0.11 and Bridge E.
The MDP values relative to each MC region in the 4$-$8 keV energy range are listed in Table \ref{tab:mdptable}. 
The MDP map confirms what found in \citet{DiGesu2020}: the cloud G0.11-0.11 has the lowest MDP, followed by Bridge E, Bridge B2, and MC2.
We note that in \citet{DiGesu2020} the MDP found for G0.11-0.11 in the 4$-$8 keV energy range in 2 Ms was 9\%, while the current value is equal to 12.5\%.
The likely reason of this difference is twofold. 
The first reason is that, as explained in Sect. \ref{sec:sourcemodel} the assumed instrumental background is higher than the model considered in \citet{DiGesu2020}.
This results in an increase of the MDP according to Eq. \ref{eq:mdp}. 
The second reason, as we discuss here below, is that in our simulations the clouds are not placed in the center of the FOV.
%
\subsection{Off-axis detectability}
\label{sec:offaxis}
\begin{figure*}[htbp]
	\centering
	\includegraphics[width=1.0\linewidth]{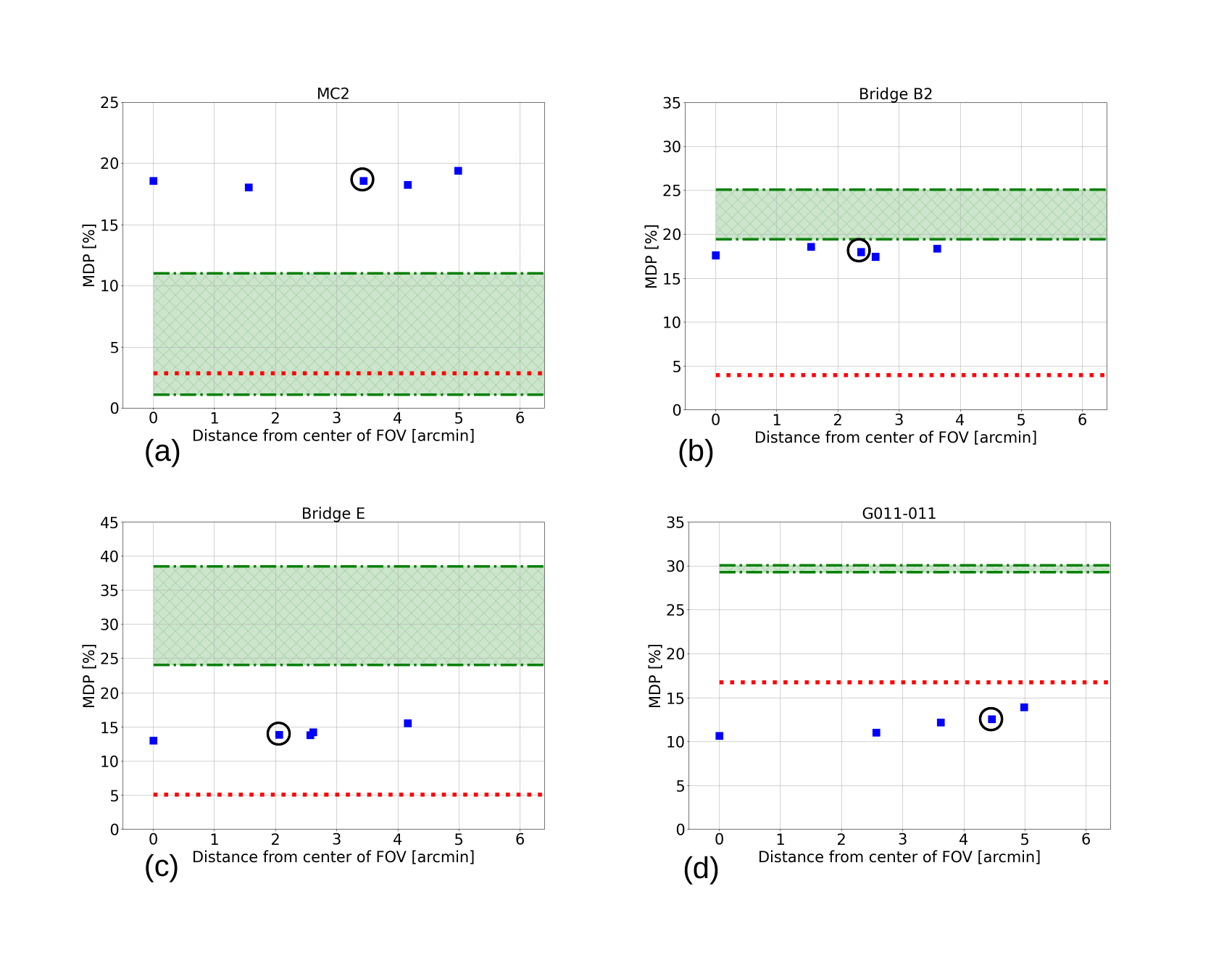} 
	\caption{
		MDP in the 4.0$-$8.0 keV band (square data points) for a 2 Ms-long observation as a function of the distance from the center of the FOV for the clouds MC2 (a), Bridge B2 (b), Bridge E (c), and G0.11-0.11 (d).
	The circled data points refer to the MC distance from the center of the FOV in the nominal IXPE pointing.
	The dashed line is the polarization degree from the model of \citet{marin2015} diluted by the unpolarized emission in the 4$-$8 keV energy range.
	The shaded region within the dash-dotted lines covers the polarization degree range predicted for other line-of-sight distances for each MC reported in Table \ref{tab:clouds}, diluted by the unpolarized emission in the 4$-$8 keV energy range.}
	\label{fig:figure_6}
\end{figure*}

We study how the MDP of the MC changes as a function of the off-axis distance. 
For this, we run simulations putting each time the clouds MC2, Bridge B2, G0.11-0.11, Bridge E, and the nominal IXPE pointing at the center of the FOV.
In Fig. \ref{fig:figure_2}, we show the regions covered by each pointing.
For each MC, we measure the MDP that can be achieved in a 2 Ms-long observation in each pointing configuration.
For this exercise, the values assumed for the polarization properties are irrelevant, as we are only interested in how the MDP changes with the distance from the center of the FOV. 
In Fig. \ref{fig:figure_6} we show for each MC the MDP as a function of the distance from the center of the FOV in the 4$-$8 keV energy band. 
We observe that the MDP of each cloud when observed at the nominal pointing increases by a factor of $\sim$1\% for MC2, $\sim$2\% and Bridge B2, $\sim$15\% for G0.11-0.11 and $\sim$6\% for Bridge E, with respect to the case of a on axis observation.
The cause of the differences in MDP across the FOV is mainly the vignetting.
The vignetting defines the relative exposure across the FOV and causes a drop of the effective area especially above 6 keV in energy and at 5 \arcmin \, in distance from the center of the FOV, thus resulting in a loss of counts for a target off-axis. 
We find that the effect of vignetting is more significant in the case of G0.11-0.11 and Bridge E, as they display a steeper increase of the MDP as a function of the off-axis distance. 
This is likely because they exhibit a harder spectrum with respect to MC2 and Bridge B2 (see Fig. \ref{fig:figure_4}) and are generally farther from the center of the FOV when the other clouds are pointed. \\
To assess the detectability of the clouds, the MDP has to compared with the expected polarization degree diluted by unpolarized ambient radiation. 
As a visual comparison, in Fig. \ref{fig:figure_6} we show also horizontal lines correspondent to the theoretical polarization degree of \citet{marin2015} and the range achievable assuming other line-of-sight distances.
These values are also listed in Table \ref{tab:mdptable}, together with the 4.0$-$8.0 keV MDP correspondent to the case of the nominal pointing. 
They differ from the ones reported in \citet{DiGesu2020} mainly because of the different background used in the present work and, at a second order, because of the updated instrumental response functions in the ixpeobssim simulator.
We find that, assuming a 2 Ms-long IXPE observation, and the \citet{marin2015} model, only the cloud G0.11-0.11 is detectable in the 4$-$8 keV energy band, even when observed off-axis.
If we assume the alternative distances of Table \ref{tab:clouds}, significant detection of polarization from the MC2 cloud remains unlikely regardless of its position in the FOV.
On the other hand, polarization degree detection at confidence level of 99\% is possible for the clouds Bridge B2 and Bridge E, as their expected diluted values of polarization degree is larger than the MDP in 2 Ms.
We note that, in case of a non detection of a cloud, the nominal MDP prescribes an upper limit to the cloud distance along the line-of-sight (Eq. \ref{eq:poldeg} and \ref{eq:distance}). 
We list these $\it \vec{d}_{\rm los}^{\rm MDP}$ in Table \ref{tab:mdptable}.
These will be valuable constraints to mitigate our uncertainty in the knowledge of the 3D position of the MC in the GC region.
%
%
\begin{table*} 
\caption{Minimum Detectable Polarization, expected diluted polarization, and $\rm |d_{\rm los}|$ relative to the MDP for the MC2, Bridge B2, G0.11-0.11, and Bridge E clouds in the 4$-$8 keV band.}     
\label{tab:mdptable}      
\centering  
{\tabulinesep=1.mm
\begin{tabu}{cccccc}
\hline \hline
\multirow{2}{*}{Cloud}		& Energy	& $\rm MDP$\tablefootmark{a}	& $\rm P_{model, diluted}$\tablefootmark{b} 	& $\rm P_{other, diluted}$\tablefootmark{c}  & $\rm |\vec{d}_{los}|^{MDP}$	\tablefootmark{d}\\
							& (keV)		& (\%)  						& (\%) 				& (\%)		& (pc)	\\
\hline
MC2							& 4-8		& 17.9							& 3 				& 1 $-$ 11	& $\geq$21 * \\
Bridge B2					& 4-8		& 17.9							& 4 				& 19 $-$ 25	& $\geq$27	\\	
G0.11-0.11					& 4-8		& 12.5							& 17				& 29 $-$ 30	& $\geq$50	\\
Bridge E					& 4-8		& 13.5							& 5					& 24 $-$ 39	& $\geq$45	\\
\hline \hline
\end{tabu}}
\tablefoot{
\tablefoottext{a}{MDP in 2 Ms in the 4.0$-$8.0 keV band correspondent to the case of the nominal pointing. Includes effect of instrumental background.}
\tablefoottext{b}{Expected polarization degree from model of \citet{marin2015} after environmental dilution.}
\tablefoottext{c}{Expected polarization degrees for other line-of-sight distances of the MC \citep{capelli2012} after dilution.}
\tablefoottext{d}{Absolute value of the line-of-sight distance that corresponds to the MDP.}
\tablefoottext{*}{For the MC2 cloud the MDP achievable in 2 Ms is not low enough to exclude the case $\it \vec{d}_{\rm los}$=0 \citep{capelli2012} because it corresponds to $\it P_{other, diluted}$=11\%, see also Fig. \ref{fig:figure_6}. Hence, for this target, the lower limit given by $\vec{d}_{\rm los}^{\rm MDP}$ in the case of a non detection cannot be taken at face value.}
}
\end{table*}
\section{Reconstruction of the intrinsic cloud polarization}
\label{sec:reconstruction}
The dilution of the polarization degree due to environmental and instrumental effects hampers the detectability of the clouds and hence the possibility to derive from the polarimetric data their line-of-sight distance.\\
We test two possible methods to create polarization products of the diluting components.
We then combine them to a test dataset simulated for a realistic observing time of 2 Ms (Fig. \ref{fig:figure_7} (a)) to test whether it is possible to reconstruct the intrinsic polarization degree of the cloud and, in turn, to derive the distance of the cloud along the line-of-sight. 
This test dataset will be replaced by real IXPE data once they will be available. \\
Both these methods are applicable only to the case of diluting components that are unpolarized. 
The case of dilution from polarized components is beyond the scope of this paper. 
As a visual comparison, we computed a map of $\it P_{\rm reflection}$ (\ref{fig:figure_7} (b)), that is the polarization degree map that IXPE would observe with no unpolarized sources in the FOV.
This was created by running a realistic simulation including only the polarized reflection continuum component. 
\subsection{Dilution map method}
\label{sec:dilution}
The first technique that we propose to reconstruct the intrinsic polarization degree of the MC consists in creating a map of the dilution factor over the entire FOV, that is a dilution map. 
Hence, the undiluted polarization degree, $\it P_{\rm dmapcorr}(x,y)$, map is obtained by dividing, pixel by pixel, the observed polarization degree map $\it P_{\rm obs}(x,y)$ by the dilution map $\it D(x,y)$: 
\begin{equation}
	\rm	P_{dmapcorr}(x,y) = \frac{P_{obs}(x,y)}{D(x,y)} \quad .
	\label{eq:P_undil}
\end{equation}
To create the dilution map, we proceeded as follows. 
We setup an ideal simulation as explained in Sect. \ref{sec:methods} and we assign a polarization degree of 100\% to all the molecular clouds and we produce a map of the Stokes parameters in the 4.0$-$8.0 keV band range.
We bin the Stokes maps so that each pixel has the size of the IXPE PSF ($\sim$30''). 
We produce a polarization map in which the polarization degree is calculated in each spatial bin from the Stokes parameters from Eq. \ref{eq:P_stokes}.
In this way, the resulting, simulated polarization map is \textit{de facto} a map of the dilution factor due to the unpolarized components.
The dilution map is shown in Fig. \ref{fig:figure_7} (c). \\
Hence, in order to test whether this technique is effective in recovering the intrinsic polarization degree of the clouds, we created the undiluted polarization map by using the formula reported in Eq. \ref{eq:P_undil}. 
This is shown in Fig. \ref{fig:figure_7} (d).
The polarization properties of individual MC are defined as the average of the values of the pixels inside the MC regions weighted by their intensity:
\begin{equation}
	\rm P_{dmapcorr}^{cloud} = \frac{\sum^{cloud} P_{dmapcorr}(x,y) I(x,y)}{\sum^{cloud} I(x,y)} \quad .
\end{equation}
These are listed in Table \ref{tab:results} together with the values of $\it D$ and $\it P_{\rm dmapcorr}$ for each cloud with their uncertainties.
The uncertainty on the value of $\it D$ is obtained from the uncertainties of the spectral fit of the Chandra data of each cloud in the following way:
\begin{equation}
\rm	\frac{\sigma_D}{D} = \frac{\sigma_{Fcrefl/Ftot}}{F_{crefl}/F_{tot}}
\label{eq:dilutionmethoderr}
\end{equation}
where $\it F_{\rm crefl}$ is the flux of the polarized reflection continuum, $\it F_{\rm tot}$ is total flux, $\it \sigma_{\rm Fcrefl/Ftot}$ is the uncertainty on the ratio of the polarized and total fluxes ($\it F_{\rm crefl}/F_{\rm tot}$).
The uncertainty on $\it P_{\rm obs}$ is obtained using equation \ref{eq:perrkislat} that includes the effect of the uncertainty in the knowledge of the subtracted background. 
Thus, the uncertainty on $\it P_{\rm dmapcorr}$ is obtained propagating the errors of $\it D$ and $\it P_{\rm obs}$. \\
The results of the dilution technique are summarized in Table \ref{tab:results}, where we report for each cloud the reconstructed intrinsic polarization degree and $\it \vec{d}_{\rm los}$, the latter calculated from Eq. \ref{eq:poldeg} and \ref{eq:distance}. 
The only target for which we are able to constrain the corrected polarization degree is G0.11-0.11, with a polarization degree of 49\% $\pm$ 20\% in the 4$-$8 keV energy band, that has to be compared to a 55.8\% polarization degree model. \\
This is expected because, in our simulations, G0.11-0.11 is the only MC for which the observed (diluted) polarization degree is larger than the MDP in the 4.0$-$8.0 keV band (see Table \ref{tab:mdptable}) and hence detectable in the first place at a confidence level of 99\% in a 2 Ms-long observation.
For the clouds MC2, Bridge B2, and Bridge E we can set upper limits to their polarization degree, and hence distances (see Table \ref{tab:results}).
The line-of-sight distance is found from Eq. \ref{eq:poldeg} and Eq. \ref{eq:distance}.
For G0.11-0.11, we obtain a distance of $\pm$19$\pm$8 pc pc in the 4$-$8 keV energy band consistent with the -17 pc of the model.
We can then break the geometric degeneracy by studying the shape at low energies of the continuum reflection \citep{churazov2002}. 
\subsection{Subtraction method}
\label{sec:subtraction}
\begin{figure*}[htbp]
	\centering
	\includegraphics[width=1.\textwidth]{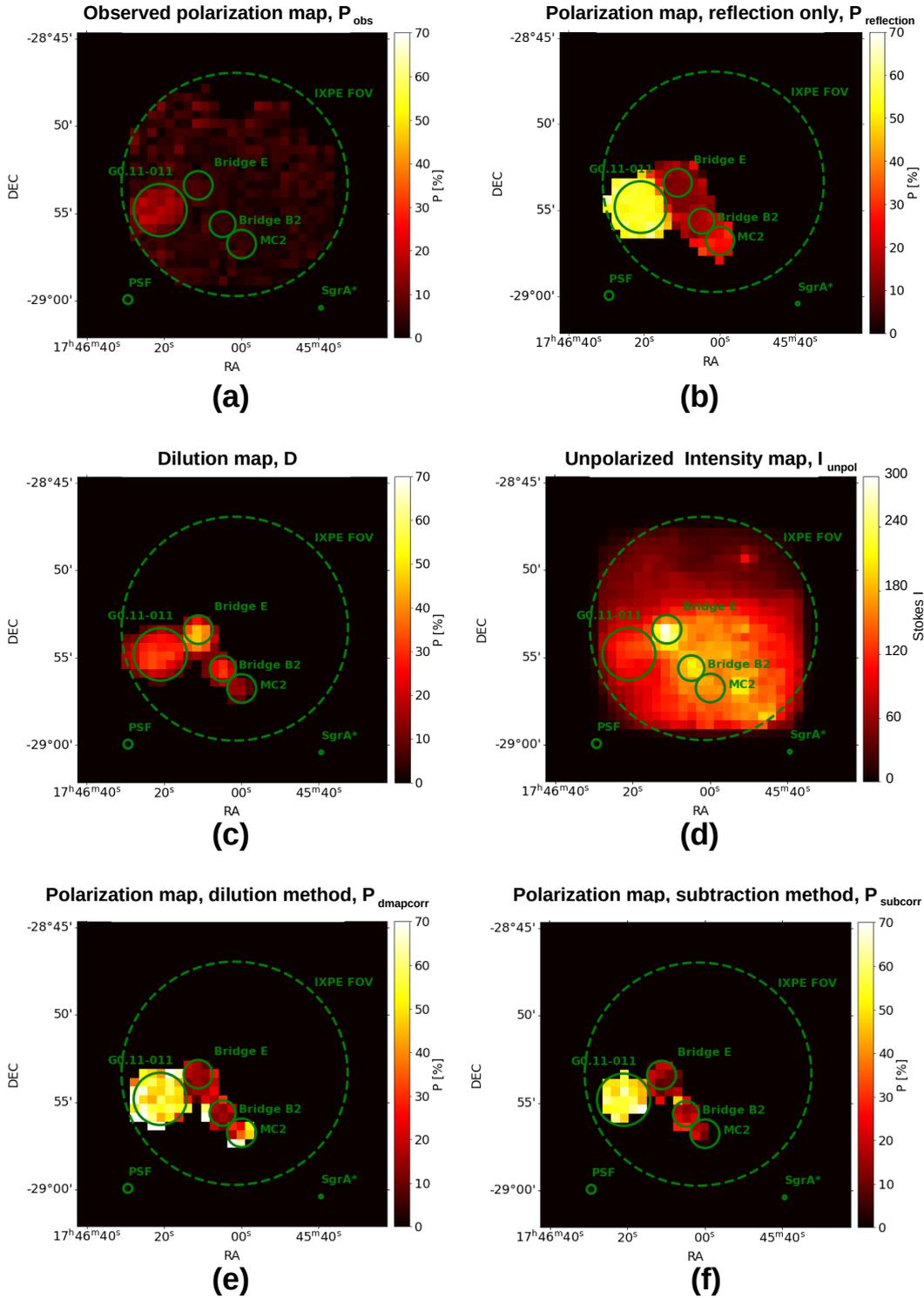}  
	\caption{
		(a) Observed polarization map. 
		For clarity, we show the results for an average of 100 simulations.
		(b) Polarization map when only the MC reflection component is considered.
		(c) Dilution map.
		(d) Unpolarized intensity Stokes parameter map. 
		(e) Reconstructed polarization map from dilution map method. 
		(f) Reconstructed polarization map from subtraction method.	
		The IXPE FOV and the cloud regions are displayed as in Fig. \ref{fig:figure_5}.	
		For the maps (a), (b), (c), (e), and (f) the color-bar displays the polarization degree, for map (d) it displays the Stokes intensity parameter.	 
		The maps have a 30 '' spatial binning and are obtained in the 4$-$8 keV energy band.}
	\label{fig:figure_7}
\end{figure*}

As an alternative, we test a second technique which exploits the additivity of the Stokes parameters. 
For an unpolarized component like the diffuse thermal emission, the $\it Q$ and $\it U$ Stokes parameters are zero.
The only relevant contribution to the dilution of the polarization degree is given by the unpolarized Intensity $\it I_{\rm unpol}$.
By subtracting from the observed Stokes Intensity map the contribution of the unpolarized emission, what remains are the Stokes parameters of the polarized component only from which the polarization degree can be computed as in Eq. \ref{eq:P_stokes}. \\
We create an intensity $\it I_{\rm unpol}(x,y)$ map of the unpolarized components, that are the soft and hard plasma, and the \fek \, line.
For this, we run an ideal simulation including the aforementioned components only.
From the simulated polarization map cube, we extract the Stokes parameter maps and we rescale them by a realistic exposure time of 2 Ms
(Fig. \ref{fig:figure_7}(d)).
Then, to mimic a real IXPE observation and create the maps of $\it I_{\rm obs}(x,y)$, $\it Q_{\rm obs}(x,y)$, and $\it U_{\rm obs}(x,y)$, we run a 2 Ms-long simulation including all the components, and with the polarization degree of the clouds set to the values of Table \ref{tab:clouds}.
The map of the intrinsic polarization degree of the MC $\it P_{\rm subcorr}(x,y)$ can be obtained using Eq. \ref{eq:P_stokes}:
\begin{equation}
\rm	P_{\rm subcorr}(x,y)=\frac{\sqrt{Q^2_{\rm obs}(x,y) + U^2_{\rm obs}(x,y)}}{I_{\rm obs}(x,y)-I_{\rm unpol}(x,y)} \quad , 
\label{eq:psub}
\end{equation}
We obtain the final $P_{\rm subcorr}(x,y)$ map (Fig. \ref{fig:figure_7}(f)), by replacing pixel-by-pixel, in the observed polarization map cubes, $I_{\rm obs}(x,y)$ with $I_{\rm obs}(x,y) - I_{\rm unpol}(x,y)$.  
The final reconstructed value of the polarization degree is the average weighted by the intensity over each cloud region (Table \ref{tab:results}). \\
We estimated the error for $\it P_{\rm subcorr}$ by propagating the error for  $\it I_{\rm obs}$, $\it Q_{\rm obs}$, $\it U_{\rm obs}$, and $\it I_{\rm unpol}$. 
We note that in our cases the Stokes parameters can be treated as independent variables because is generally true that $P \mu < 0.3$ \citep{Kislat2015}.
The uncertainty on all the observed Stokes parameters are an output of the realistic simulation and include the uncertainty in the knowledge of the subtracted background. 
The uncertainty on $\it I_{\rm unpol}$ derives from the uncertainties of the fits of the Chandra data in the following way:
\begin{equation}
\rm 	\frac{\sigma_{Iunpol}}{I_{unpol}} = \frac{\sigma_{(Fsoft\,plasma + Fhard\,plasma + FK\alpha)}}{F_{soft\,plasma} + F_{hard\,plasma} + F_{K\alpha}} \quad ,
\label{eq:subtractionmethoderr}
\end{equation}
where $\it F_{\rm soft\,plasma}$, $\it F_{\rm hard\,plasma}$, and $\it F_{\rm K\alpha}$ are the fluxes of the soft plasma, the hard plasma, and the \fek \, line, respectively. 
We note that it is critical to determine correctly the \fek\, contribution because, as shown in Table \ref{tab:clouds}, its flux is always one order of magnitude larger than the continuum. \\
We checked that in 2 Ms the contribution to $\it Q_{\rm obs}$ and $\it U_{\rm obs}$ of the random fluctuation of the unpolarized component is smaller than the uncertainty on $\it Q_{\rm obs}$ and $\it U_{\rm obs}$. \\
The reconstructed polarization degree map obtained with this method is shown in Fig. \ref{fig:figure_7} (f) while the polarization degree averaged in each cloud region are listed in Table \ref{tab:results}.\\
We find that this procedure returns a constrained value for G0.11-0.11 in the 4$-$8 keV energy band, with a reconstructed polarization degree of 53\% $\pm$ 13 \%.
We calculated the $\it \vec{d}_{\rm los}$ resulting from the reconstructed polarization with this method, and we list their values in Table \ref{tab:results}.
For G0.11-0.11 we obtain a distance of $\pm$18 $\pm$ 4 pc in the 4$-$8 keV energy band, consistent with the -17 pc assumed in the model. 
Again, the geometrical degeneracy can be removed thanks to the shape of the reflection continuum \citep{churazov2002}. \\
%
%
\begin{table*} 
\caption{Results of the reconstruction of the polarization degree of the MC of the SgrA complex in the 4$-$8 keV energy band with the dilution and subtraction methods.}     
\label{tab:results}      
\centering  
{\tabulinesep=1.mm
\begin{tabu}{cccccccc}
\hline \hline
\multirow{2}{*}{Cloud}		& Energy	& $\rm P_{obs}$\tablefootmark{a} &  $\rm D$\tablefootmark{b} 	& $\rm P_{dmapcorr}$\tablefootmark{c} 	& $\rm P_{subcorr}$\tablefootmark{d} & $\rm |\vec{d}_{los}|^{dmapcorr}$	\tablefootmark{e}	& $\rm |\vec{d}_{los}|^{subcorr}$	\tablefootmark{f}  	 \\
							& (keV)		& (\%)			& (\%) 			& (\%)			& (\%)			& (pc)			& (pc) \\
\hline
MC2							& 4-8		& $\leq$17.9	& 11 $\pm$ 9	& $\leq$53		& $\leq$30		& $\geq$9	& $\geq$15 	\\
\hline
Bridge B2					& 4-8		& $\leq$17.9	& 25 $\pm$ 6	& $\leq$45		& $\leq$34		& $\geq$13	& $\geq$18  	\\
\hline
G0.11-0.11					& 4-8		& 15 $\pm$ 6	& 30 $\pm$ 3	& 49 $\pm$ 20	& 53 $\pm$ 13  	& 19$\pm$7	& 18$\pm$4  	\\
\hline
Bridge E					& 4-8		& $\leq$13.5	& 39 $\pm$ 4	& $\leq$31		& $\leq$20  	& $\geq$26	& $\geq$35 	\\
\hline \hline
\end{tabu}}
\tablefoot{
\tablefoottext{a}{Observed polarization degree with a 2 Ms-long observation.
	The uncertainties on the observed polarization degree include the effect of the subtraction of the instrumental background.}
\tablefoottext{b}{Dilution factor obtained from the dilution map described in Sect. \ref{sec:dilution}.}
\tablefoottext{c}{Polarization degree recovered through the dilution method.}
\tablefoottext{d}{Polarization degree recovered through the subtraction method.}
\tablefoottext{e}{Absolute value of the line-of-sight distance derived from the polarization degree corrected with the dilution method.}
\tablefoottext{f}{Absolute value of the line-of-sight distance derived from the polarization degree corrected with the subtraction method.}
}
\end{table*}
%
%
\section{Discussion}
\label{sec:discussion}
In this work, we estimated the MDP reached in a single 2 Ms observation with IXPE of the MC of the Sgr A complex in the 4$-$8 keV energy band.
Our estimations consider two additional factors that we did not take into account in our previous work \citep{DiGesu2020}.
The first one is the effect of the vignetting of the telescope optics, that causes a loss of counts, and hence an increase of the MDP. 
The second one is the updated background model, as derived by \citet{xie2021}, for the IXPE detectors that is larger by a factor of three than the one based on \citet{bunner1987}. 
The scientific case considered here in one of the few for which the instrumental background is a potential confounding factor because  of the faintness of the MC. 
\\
We found that the MDP of the MC obtained in the case of the nominal IXPE pointing of the region, because of vignetting, increases by a factor in the range of 1$-$15\% with respect to the case in which each of them is observed on-axis. 
Vignetting effects will be negligible for most sources that IXPE will observe, because the telescope pointing will be dithered around the center of the FOV.
The observation of the MC in the GC is one of the few cases in which vignetting will have an observable effect.
\\
Assuming the model of \citet{marin2015}, in a 2 Ms-long IXPE observation, G0.11-0.11 is the only MC detectable.
However, the prediction of the polarization degree depends strongly on the assumed line-of-sight distance. 
When changing the assumption on $\it \vec{d}_{\rm los}$ (Table \ref{tab:clouds}), for MC2, Bridge B2, and Bridge E we find a higher polarization degree.
In this alternative scenario, the latter two clouds are detectable in a 2 Ms-long observation for all the pointings considered in this work. 
MC2 is undetectable in a 2 Ms-long observation for any polarization model. 
This is because the cloud is the faintest among the ones we considered in this work, and it is affected by the worst environmental dilution, $\sim90\%$ (see Table \ref{tab:clouds}).
\\
With the assumed distances (and hence polarization degree, see Eq. \ref{eq:poldeg}, \ref{eq:distance}) of the clouds, only G0.11-0.11 appears to be a candidate for a statistically significant measurement of the X-ray polarization in the Sgr A complex. 
The possibility of recovering the polarization degree depends on the a-priori significativity of the measurement of the diluted polarization degree. \\
In this work we tested two methods to recover the intrinsic polarization degree of the MC in the 4$-$8 keV energy range, where the polarized reflection outshines the unpolarized plasma emission. 
The dilution map method, described in Sect. \ref{sec:dilution}, consists in dividing pixel-by-pixel two polarization maps: the observed polarization map and a dilution map.
We created the dilution map by simulating the case of clouds 100\% polarized. 
In this way, the resulting polarization degree image maps the dilution factor over the FOV.
This method allows to remove the depolarizing effect of the plasma and the emission of the \fek \, line.
For G0.11-0.11, from a polarization degree of 15 $\pm$ 6\%, the dilution method recovers a value of 49 $\pm$ 20\%, consistent within the uncertainty with the input model of 55.8\%. 
For G0.11-0.11, this method allows to recover the line-of-sight distance of the cloud with Eq. \ref{eq:poldeg} and \ref{eq:distance} as $\pm$19$\pm$7 pc, consistent with the -17 pc of the model. \\
The subtraction method, described in Sect. \ref{sec:subtraction}, is based on the subtraction from the observed Stokes I parameters map of the Stokes maps of the unpolarized components only. 
The residual parameters are employed for the calculation of the polarization degree. 
For G0.11-0.11, the subtraction methods gives a polarization degree value of 53 $\pm$ 13\%.
Again the reconstructed line-of-sight distance of $\pm$18$\pm$4 pc is consistent with the input model. 
In both cases, the ambiguity on the position can be removed through spectroscopic means by studying the shape of the reflection continuum, as explained in \citet{churazov2002}: if the cloud is closer to the observer with respect to the illuminating source, the reprocessed radiation from the farthest, directly illuminated, side of the cloud would be suppressed at low energies by photo-absorption. \\
Besides the uncertainty in the cloud distance, and hence in the theoretical polarization degree, there are other potential challenges for the planning of an IXPE observation of the GC. 
The MC exhibit a time variability in flux and morphology on $\sim$years timescale. 
For instance, \citet{terrier2018} note that the flux of the clouds Sgr B2 and G0.74-0.10 decreased by a factor factor 4$-$5 over 12 years. 
In the case of G0.11-0.11, the brightest \fek\, feature shifted towards the Galactic East by $\sim$3 arcmin in 12 years. \\
Hence, it is fundamental that the IXPE observation is complemented by a quasi simultaneous pointing with another X-ray facility that provides the up-to-date morphology and spectrum of the clouds. 
Using IXPE maps only, it is difficult to pinpoint the location of the brightest \fek\, patches. 
The spectral resolution of IXPE at 6 keV is $\sim$1 keV, thus the \fek\, line from the clouds is blended with the \ion{Fe}{xxv}-He$\alpha$ and \ion{Fe}{xxvi}-Lyman lines of the hard plasma. 
This means that, using IXPE data only, we may be not able to identify the optimal regions where the reflection of the clouds prevails over the plasma emission.
The up-to-date \fek\ morphology of the Sgr A region can be provided either by Chandra, XMM-Newton, or eROSITA. 
However, only Chandra maps can be input in our procedures to compute the synthetic maps of the unpolarized components. 
This is because the resolution of Chandra is infinite from the IXPE point of view. 
For this reason, a Chandra observation is the best complement to the IXPE observation and would allow to treat the data exactly as we outlined in this paper.
In the absence of a simultaneous Chandra coverage, it would still be feasible to apply our correction methods using Chandra archival maps and spectra for the plasma components and complement them with an up-to-date spectrum of the clouds provided by e.g. XMM-Newton or eROSITA, as the latter performs a monitoring of the GC every six months. 
However, in this case, a uniform morphology must be assumed for the cloud while computing the synthetic maps the unpolarized components. 
We already checked in \citet{DiGesu2020} that simulating the clouds as a uniform source does not change the results. \\
Without an up-to-date spectrum of the clouds, our methods cannot be applied, and the intrinsic polarization of the clouds cannot be retrieved correctly. 
The better the quality of the spectrum, the better the final uncertainties on the measurement of the distance along the line-of-sight of the cloud.
In order to quantify this point we make the exercise of checking how much the uncertainty on both our methods would change if the knowledge of the up-to-date spectrum comes from IXPE data only. 
We fit a simulated IXPE spectrum of G0.11-0.11 with an absorbed APEC+APEC+PEXMON model.
We computed the errors for $\it F_{\rm crefl}$, $\it F_{\rm tot}$, $\it F_{\rm K\alpha}$ and we used Eq. \ref{eq:dilutionmethoderr} and \ref{eq:subtractionmethoderr} to evaluate the uncertainties on the reconstructed polarization. 
We find that the uncertainty on the reconstructed polarization worsens by a relative factor of 20\% and 46\% with the dilution and subtraction methods, respectively. 
The uncertainty on the line-of-sight distances increases to $19_{-8}^{+14}$ pc and $18_{-6}^{+9}$ pc with the dilution and subtraction methods, respectively. 
Thus, we expect that the spectral quality of Chandra, XMM-Newton or eROSITA will ensure that the distance of the cloud along the line-of-sight is determined with an uncertainty of the order of a few parsec. \\
Finally we note that, in case a strong variability in either flux or morphology of the clouds is detected before the IXPE pointing, the MDP of the clouds must be updated to decide the best pointing strategy. 
Recomputing the MDP with new input spectra and cloud location is straightforward using our procedure.\\
From the constrains on the polarization degree depends the possibility of reconstructing the 3D distribution of the gas in the CMZ.
Even in case of non detection, an X-ray polarimetric study of the MC will put useful constraints on their position along the line-of-sight. 
These values will be determined by the nominal MDP in the cloud region at the time of the IXPE observation. 
As an example, we made the exercise of computing $\it |\vec{d}_{\rm los}|^{\rm MDP}$ for all the clouds considered here (see Table \ref{tab:mdptable}). 
We note that these would be model-independent constraints because only the number of counts collected during the observation is needed to determine the MDP. 
The other method currently available to derive the line-of-sight \citep[e.g.][]{capelli2012} rely on the measurement of the equivalent width of the \fek \, line, that depends on the scattering angle because of the angular dependence of the scattering continuum. 
This require a careful modeling of the reflection continuum and an accurate knowledge of the Iron abundances in the GC region. \\
We note that the uncertainty on the reconstructed value of the polarization degree is always slightly larger with the dilution method with respect to the uncertainty of the subtraction method.
This is because the uncertainty on the dilution method depends on the dilution factor $\it D$ that, in our model, is no larger than the 39$\pm$4\% estimated in the Bridge E region. 
Even in the case in which the polarization is in principle undetected, this results in different $\it d_{\rm los}$ estimates because of the different environmental dilution in each cloud region.
For this reason, the subtraction methods returns more accurate results.\\ 
All in all, the capability of both our methods to recover the intrinsic polarization properties of the clouds is supported by the results described in Sect. \ref{sec:reconstruction}.
The comparison between the undiluted polarization map shown in Fig. \ref{fig:figure_7} (b) and the reconstructed polarization maps with the dilution and subtraction methods shown in Fig. \ref{fig:figure_7} (e) and \ref{fig:figure_7} (f), respectively, visually highlights the efficiency of our methods in cleaning up the data from the contamination of the unpolarized emission. 
Both methods presented here to recover the intrinsic X-ray polarization degree are not limited to IXPE but could also be employed to treat the data coming from future X-ray polarimetry missions such as the enhanced X-ray Timing and Polarimetry mission \citep[eXTP,][]{extp}, the Next Generation X-ray Polarimeter \citep[NGXP,][]{esa2050}, or the X-ray Polarization Probe \citep[XPP,][]{xpp}. 
These missions will have even greater sensitivity and spatial resolution with respect to IXPE. 
This highlights the importance of having tested general methods including detailed morphological information, allowing in the future to produce synthetic products suitable for instruments with any angular resolution. 
Moreover, the methods are relevant not only for the GC case, but also for all the cases where the expected polarization signal from extended sources is contaminated by an unpolarized diffuse emission.
An example of such extended sources are the supernova remnants (SNR) Cas A and Tycho. 
In these sources, the X-ray synchrotron emission, that is expected to be highly polarized \citep[e.g.][]{Bykov2017}, is mixed with an unpolarized multi-temperature plasma emission.
Chandra maps and spectra of the unpolarized components of the named SNR could be fed into in the procedures described here to recover the intrinsic polarization degree. 
%
\section{Summary and conclusions}
\label{sec:conclusion}
Spatially-resolved X-ray polarization measurements of the molecular clouds (MC) in the Galactic center would allow us to test the hypothesis that they are reflecting X-rays from a past outburst of the now under luminous SMBH Sgr A*. 
However, contamination from unpolarized sources, instrumental background, low luminosity, and extension of the clouds, make this experiment a challenging one. 
In this paper we described data analysis techniques for the upcoming measurement of X-ray polarization from the MC in the Sgr A complex with the Imaging X-ray Polarimetry Explorer (IXPE). 
The launch of IXPE is expected in late 2021, allowing these techniques to be tested on real data. \\
We simulated a 2 Ms-long IXPE observation of the Sgr A region. 
The IXPE field of view (FOV) includes four molecular clouds (MC2, Bridge B2, Bridge E, and G0.11-0.11) that are embedded in the diffuse plasma of the GC region. 
We used Chandra maps and spectra to model the spectrum and the morphology of the clouds and of the diffuse unpolarized thermal emission. 
This unpolarized emission has the effect of diluting the polarized signal. 
We also included the CXB and the instrumental background. \\
We produced a map of the Minimum Detectable Polarization (MDP) that can be obtained with a 2 Ms-long exposure of the Sgr A MC complex in the 4$-$8 keV energy band.
We evaluated the effect on the reduction of the MDP due to the fact that the clouds are observed off-axis in the FOV.
We demonstrated that the MDP, with respect to an on-axis observation of each cloud, increases by a factor $\sim$ 1$-$15\% due to vignetting effects depending on the cloud position in the FOV and spectral shape. 
In addition, we presented two independent techniques to recover the intrinsic polarization degree of the MC.
We demonstrated that these two techniques can recover the polarization degree, and hence the line-of-sight distance $\it \vec{d}_{\rm los}$ of a cloud whose polarization is detected at a 99\% confidence level. 
For instance, for G0.11-0.11 we find that for a 2 Ms-long IXPE observation in the 4$-$8 we can constrain the distance along the line-of-sight with respect to the Galactic plane to $\pm$19$\pm$7 pc and $\pm$18$\pm$4 pc with the dilution and subtraction method, respectively. \\
Because the brightness of these clouds changes with time, a Chandra observations quasi-simultaneous with the IXPE targeting of the Galactic center will assess the illumination status of the clouds, and provides the morphological and spectral information needed to apply the polarization recovery methods described.
We estimate that, with this observation strategy, the uncertainty on the measurement of the line-of-sight when a cloud is detected will be of the order of a few parsec.
The same approach can be applied to future X-ray polarimetric missions such as eXTP, NGXP, and XPP, and to other extended sources where the polarized signal may be diluted by unpolarized diffuse and mixed components, such as in supernova remnants. 
%
%
%

\bibliography{merged}
\bibliographystyle{aa}

%

\begin{acknowledgements}
The Italian contribution to the IXPE mission is supported by the Italian Space Agency through agreements ASI-INAF n.2017-12-H.0 and ASI-INFN n.2017.13-H.0.
FM acknowledges the support from the Programme National des Hautes Energies of CNRS-INSU with INP and IN2P3, co-funded by CEA and CNES.
We thank Sergio Fabiani for the useful chat about the uncertainties of the methods.
We thank the anonymous referee for the helpful comments that improved this manuscript.
\end{acknowledgements}

\appendix

\begin{appendices}
\section{Instrumental background effect and treatment}
\label{sec:appendix}
We consider the realistic simulations as defined in \ref{sec:simulation} as instrumental background and CXB-subtracted.
When actual IXPE data will be available, it will be possible to subtract from the observation the Stokes parameters of the instrumental background.
Information on the IXPE instrumental background will be collected during the mission lifetime through measurements taken during Earth-occultations, with the black filter present in the on-board filter and calibration wheel covering the detector sensitive area \citep{Ferrazzoli2020JATIS}.
Finally, it will be possible to estimate the background from the unused FOV when observing high Galactic latitude point sources. \\
The subtraction of the background Stokes parameters introduces an uncertainty on the determination of the reconstructed polarization fraction, that is given by \citet{Kislat2015} as
\begin{equation}
	\rm	\sigma_{P} = \sqrt{\frac{\rho_{BS} (2 - P^2 \mu^2)}{[\rho_{BS} (2 R_B + R_S) - 2(R_B^2 + RB R_S)] T \mu^2}}
	\label{eq:perrkislat}
\end{equation}
where $\it \rho_{BS} = \sqrt{R_B (R_B + R_S)}$, $\it P$ is the polarization degree estimated after the background subtraction, and the other parameters as in Eq. \ref{eq:mdp}. \\
This is the uncertainty that is associated with the observed polarization degree in Table \ref{tab:results}. \\
We assume that the instrumental background is known for an exposure of 2 Ms.
For the expected instrumental background rate on the G0.11-0.11 region, this amounts to an uncertainty on the determination of the instrumental background rate of
\begin{equation}
\rm \frac{\sigma(R_B)}{R_B} = \frac{1}{\sqrt{R_B T}} \approx \rm 1.9 \%
\end{equation}
with instrumental background counting rate $\it R_B = \rm 1.2 \times 10^{-3}$ c/s in the 4$-$8 keV energy band and observation time $\it T = \rm 2 \times 10^6$ s. \\
For the same cloud and exposure time, the unpolarized components counting rate $\rm 2.3 \times 10^{-3}$ c/s is known with an uncertainty of $\sim$1.5\%. 
\end{appendices}
	
\end{document}